\documentclass[preprint]{aastex}

\shorttitle{Reconnection Rate in Solar Flares}
\shortauthors{Isobe et al.}

\begin{document}


\title{Measurement of the Energy Release Rate and 
the Reconnection Rate in Solar Flares}


\author{Hiroaki Isobe\altaffilmark{1}, 
Hiroyuki Takasaki, and Kazunari Shibata}
\affil{Kwasan and Hida Observatories, Kyoto University, 
Yamashina, Kyoto 607-8471, Japan}
\email{isobe@kwasan.kyoto-u.ac.jp}

\altaffiltext{1}{present address: Department of Earth and 
Planetary Science, University of Tokyo, 
Hongo, Bunkyo-ku, Tokyo 113-0033, Japan}



\begin{abstract}
By using the method presented by Isobe et al. (2002), 
the non-dimensional reconnection rate $v_{in}/v_A$ has been 
determined for the impulsive phase of three two-ribbon flares, 
where $v_{in}$ is the velocity of 
the reconnection inflow and $v_A$ is the Alfv\'{e}n velocity. 
The non-dimensional reconnection rate is important to 
make a constraint on the theoretical models of magnetic reconnection. 
In order to reduce the uncertainty of the reconnection rate, 
it is important to determine the energy release rate of 
the flares from observational data as accurately as possible. 
To this end,  we have carried out one dimensional hydrodynamic 
simulations of a flare loop and synthesized the count rate detected by  
the soft X-ray telescope (SXT) aboard Yohkoh satellite. 
We found that the time derivative of the thermal energy contents 
in a flare arcade derived from SXT data is smaller than 
the real energy release rate by a factor of 0.3 -- 0.8, 
depending on the loop length and the energy release rate. 
The result of simulation is presented in the paper and 
used to calculate the reconnection rate. 
We found that reconnection rate is 0.047 for the X2.3 flare 
on 2000 November 24, 0.015 for the M3.7 flare on 
2000 July 14, and 0.071 for the C8.9 flare on 2000 November 16. 
These values are similar to that derived from 
the direct observation of the reconnection inflow by \citet{yok01}, 
and consistent with the fast reconnection models such as 
that of \citet{pet64}. 
\end{abstract}



\keywords{Sun: flares --- Sun: X-rays, gamma rays --- 
Sun: magnetic fields}


\section{INTRODUCTION}

Magnetic reconnection is now widely believed to be 
the mechanism of energy release in solar flares. 
The evidence of magnetic reconnection found by space 
observations includes the cusp-shaped post flare loop \citep{tsu92}, 
the loop top hard X-ray source \citep{mas94}, 
the reconnection inflow \citep{yok01},  
downflows above post flare loops \citep{mck99, inn03, asa04a}, 
plasmoid ejections (Shibata et al. 1995; Ohyama \& Shibata 1997, 1998), etc 
(see reviews by Shibata 1999; Martens 2003). 
Magnetic reconnection also plays an important role in various explosive 
phenomena in astrophysical, space, and laboratory plasmas 
\citep{bis92, taj97, pri00}. 

One important question in the physics of reconnection is 
what is and what determines the reconnection rate in a plasma 
with large magnetic Reynolds number such as solar corona. 
The reconnection rate is the amount of reconnected magnetic flux 
per unit time and given by $v_{in}B$, where $v_{in}$ 
is the velocity of reconnection inflow into the diffusion region and 
$B$ is the magnetic field strength. It is more convenient and physically 
meaningful to define it in non-dimensional form as $M_A = v_{in}/v_A$, 
where $v_A$ is the Alfv\'{e}n velocity. In the following we call 
this non-dimensional value the reconnection rate. 
It is important to determine the non-dimensional reconnection rate  
because theories of magnetic reconnection predict 
the non-dimensional value. 

In classical Sweet-Parker model 
\citep{par57, swe58} the reconnection rate $M_A$ is given by 
\begin{equation}
M_A = R_m^{-1/2}, 
\end{equation}
where $R_m$ is the Magnetic Reynolds number defined by the 
Alfv\'{e}n velocity (also called the Lundquist number): 
\begin{equation}
R_m = \big( \frac{v_A L}{\eta} \big), 
\end{equation} 
$L$ is the length of the reconnecting current sheet and 
$\eta$ is the magnetic diffusivity.  
If one consider the classical resistivity by Coulomb collision 
\citep{spi56}, the Magnetic Reynolds number is enormously large 
($\sim 10^{14}$) in the solar corona. 
Therefore the reconnection rate is too small 
to explain the rapid energy release of solar flares, where 
the energy release occurs in 10--100 $\tau_A$ ($=L/V_A$;  
Alfv\'{e}n time). 

\citet{pet64} considered the effect of MHD slow shocks and proposed 
a model in which the reconnection rate is about 0.01--0.1 
and almost independent of the Magnetic Reynolds number 
($M_A \propto [\ln R_m]^{-1}$). Several 
MHD simulations have shown that localized resistivity such as anomalous 
resistivity leads to Petschek type fast reconnection (e.g.,  
\citealt{uga86, sch89, yok94}).  
\citet{erk00} also showed that the Petschek regime was realized 
only when the resistivity is localized, by matching the Petschek-like 
solution in the inflow region and internal diffusion region 
solutions. 
\citet{lit96} examined the reconnection rate in flux-pile-up reconnection 
and suggested that the reconnection rate is slower than 
the value of Petschek reconnection by a factor of plasma beta. 
\citet{sni04} considered the self-similar solution of Petschek type 
reconnection (i.e., with slow shocks) 
in free space and found that the reconnection 
rate is $\sim 0.05$, nearly independent of the plasma beta and, 
possibly, of the magnetic Reynolds number, too. 
It is also worth noting that signature of slow mode MHD shocks associated 
with reconnection was found recently by X-ray observation of a giant 
arcade formation event \citep{shd03}. 

Observational measurement of reconnection rate is not straightforward 
because measurement of coronal magnetic field strength and 
inflow velocity is difficult. The only two direct observations of 
reconnection inflow reported in the literature are those of 
\citet{yok01} and \citet{lin05}. 
\citet{yok01} found in EUV images two clear patterns 
approaching toward the reconnection X-point with a velocity about 
5 km s$^{-1}$. The estimated reconnection rate was 0.001-0.03. 
\citet{lin05} also found a signature of flows toward reconnecting 
current sheet in a CME-related event and obtained the reconnection 
rate of 0.01-0.23.

Observations of the chromosphere provide us alternative ways 
to infer the reconnection rate in the corona. 
When magnetic reconnection occurs 
in the corona, the released energy is transported along 
the magnetic field to the chromosphere by nonthermal high energy 
particles and/or heat conduction, brightening the footpoints of 
the reconnected field lines. Therefore successive brightenings 
in the chromosphere maps the successive 
magnetic reconnection in the corona. In the standard 
two-dimensional geometry, the successive reconnection of outer 
field lines results in the separation of two ribbons 
\citep{for00, lin04}, and the (dimensional) reconnection rate, 
that is also equal to the electric field in the reconnecting 
current sheet, is given by
\begin{equation}
v_{in}B_{cr} = E_{cs} = v_{foot}B_{foot}. 
\label{eq:vinb}
\end{equation}
Here $B_{cr}$ and $B_{foot}$ are the magnetic field strength in the corona 
and at the footpoint, respectively, $v_{foot}$ is the separation velocity 
of the footpoints (two ribbons), $E_{cs}$ is the electric field in the 
current sheet, and $v_{in}B_{cr}$ is the dimensional reconnection rate. 
Several authors derived $E_{cs}$ (or dimensional reconnection rate) 
using the observation of chromospheric two ribbons and photospheric 
magnetograms (e.g., Qiu et al. 2002, 2004, Wang et al. 2003, 
Fletcher, Pollock \& Potts 2003, Asai et al. 2004b). 

Isobe et al. (2002) considered equation (\ref{eq:vinb}) and 
the energy release rate to derive the inflow velocity, magnetic 
field strength in the corona, and hence non-dimensional 
reconnection rate. They used soft X-ray images taken by
the Soft X-ray Telescope (SXT; Tsuneta et al. 1991) 
aboard {\it Yohkoh} satellite to derive the energy release rate 
and $v_{foot}$, and obtained $M_A \approx 0.001 - 0.01$ 
for the decay phase of a long duration flare. 

\begin{figure}
\epsscale{.80}
\plotone{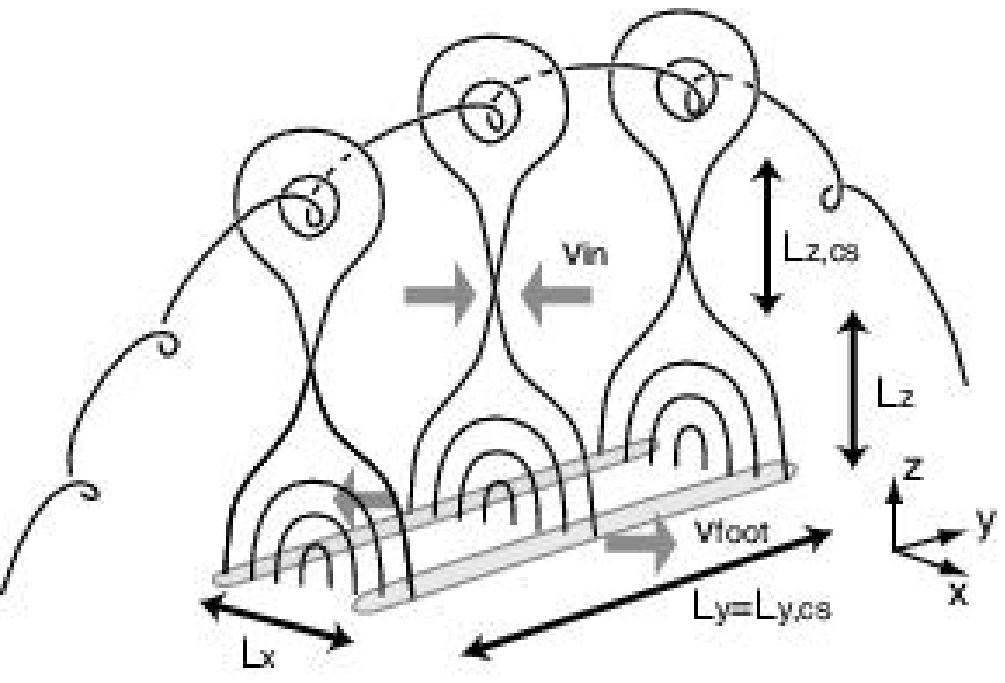}
\caption{Schematic illustration of the reconnection model of a two 
ribbon flare. 
\label{fig:model}}
\end{figure}

In this paper we apply the method presented by Isobe et al. (2002) 
to the impulsive 
phase of two ribbon flares and determine the reconnection rate $M_A$.  
We analyze three flares with different X-ray flux (one X-class, 
one M-class, and one C-class) to see whether there is a dependence 
of $M_A$ on the flare class. 
We selected the flares that exhibit relatively symmetric and 
regular separation of the flare ribbons, because 
equation (\ref{eq:vinb}) is valid  
only in the approximately two-dimensional geometry 
such as shown in figure \ref{fig:model}. 
We should note that 
it is possible that the successive brightening of the chromosphere 
is due to the successive magnetic reconnection along the magnetic 
neutral line, i.e., $y$ direction in figure \ref{fig:model}. 
In this case the velocity of the motion of chromospheric brightening 
cannot be used to calculate the electric field in the corona. 
The high electric field (90 V cm$^{-1}$) in a complicated 
C9.0 class flare reported by \citet{qiu02} may be due to this effect. 
Indeed systematic successive reconnection along the neutral line 
has been observed in giant arcade formation events, where 
the formation of X-ray loop progress along the neutral line 
with an apparent velocity of the order of 10 km s$^{-1}$
\citep{iso02g}. Interestingly, similar but faster 
(50 - 150 km s$^{-1}$) evolution of reconnection along neutral 
line has been reported recently in a M-class flare observed 
by RHESSI \citep{gri05}. 

In order to calculate the energy release rate from SXT data as accurately 
as possible, we have carried out numerical simulations of a flare 
loop. We synthesized the count rates of different filters of SXT from 
the simulation result and calculate the thermal energy contents $E_{th}$ 
that {\it will be observed} by the SXT, thus enabling us to convert 
the time derivative of $E_{th}$ obtained from SXT data to actual 
energy release rate. 

In section 2, the method to determine the reconnection rate is briefly 
explained. In section 3, we describe the model and results of 
numerical simulations. In section 4 we describe the observation, 
data analyses, and results. Discussion and conclusions are given 
in section 5.

\section{HOW TO DETERMINE THE RECONNECTION RATE} \label{sec:howto}

In this section we briefly summarize the procedure to calculate the 
reconnection rate, which is basically the same as that in Isobe et al. 
(2002) but uses different data set. 

We consider the energy release rate 
$H$ by magnetic reconnection, that is given by the Poynting 
flux into the diffusion region; 
\begin{equation}
H=2\frac{B_{cr}^2}{4\pi}v_{in}L_{y,cs}L_{z,cs}, 
\label{eq:h}
\end{equation}
where $L_{z,cs}$ is the vertical length of the reconnecting current 
sheet, and $L_{y,cs}$ is the length of reconnecting current sheet 
along the magnetic neutral line. The geometry is illustrated in 
figure \ref{fig:model}. 
The basic idea is that in equations 
(\ref{eq:vinb}) and (\ref{eq:h}) the unknown parameters are 
$B_{cr}$ and $v_{in}$ and the other parameters are obtained 
from observational data.

As mentioned in the previous section, 
$v_{foot}$ can be measured from a time sequence of chromospheric 
images of two ribbons, and $B_{foot}$ can be measured from 
photospheric magnetograms. In this paper we use 
1600 \AA \, images from the {\it Transition Region and Coronal Explorer} 
(TRACE; \citealt{han99}) to measure $v_{foot}$, and 
magnetograms from Michelson Doppler Imager (MDI; \citealt{sch95}) 
aboard the {\it Solar and Heliospheric Observatory} 
(SOHO; \citealt{dom95}) to measure $B_{foot}$. 

Since the reconnecting current 
sheet is not usually visible, $L_{y,cs}$ and $L_{z,cs}$ are 
uncertain. We assume $L_{y,cs}$ is equal to the length of 
flare arcade $L_y$. For $L_{z,cs}$, it seems reasonable to 
assume that in the impulsive phase $L_{z,cs}$ is equals to the 
height of flare arcade $L_z$. This assumption is consistent with 
MHD simulations \citep{cpf00, shd03} and the inflow observation 
\citep{yok01}. In the decay phase $L_{z,cs}$ may be much larger than $L_z$. 
In this paper we analyze flares near the disk center, therefore 
we make further assumption that $L_z$ is equal to the distance 
of the footpoints (flare ribbons) $L_x$. It is not a bad assumption 
since many flare loop near the solar limb shows such geometry. 
Namely, we assume $L_{z,cs}=L_z=L_x$. 

To measure the energy release rate $H$, we use the SXT data. 
The temperature $T$ and volume emission measure $\rm{\epsilon_V}=n^2V$ 
of the flare 
plasma can be derived from a pair of SXT images with different 
filters \citep{har92}. Then density $n$ and the thermal energy 
$E_{th}$ can be calculated by assuming a suitable line-of-sight depth. 
The energy release rate $H$ can be inferred from the time derivative 
of the thermal energy $dE_{th}/dt$, although the effect of 
radiative cooling, conductive cooling and chromospheric evaporation 
must be considered. In the next section we present the results of 
numerical simulations to convert the observed $dE_{th}/dt$ to the 
actual energy release rate $H$. 

Finally we calculate the unknown parameters $B_{cr}$ and $v_{in}$ 
from equations (\ref{eq:vinb}) and (\ref{eq:h}). The Alfv\'{e}n 
velocity $v_A = B_{cr}/\sqrt{4 \pi m_p n_0}$ ($m_p$: proton mass) 
is calculated using the calculated value of $B_{cr}$ and 
density $n_0$ in the pre-flare phase derived from the 
SXT images. Thus we obtain the non-dimensional reconnection rate 
$M_A = v_{in}/v_A$. To summarize, $B_{cr}, v_{in},$ and $M_A$ 
are given by 
\begin{equation}
B_{cr}=\frac{2\pi H}{v_{foot}B_{foot}L_xL_y}, 
\label{eq:bc}
\end{equation}
\begin{equation}
v_{in}=\frac{(v_{foot}B_{foot})^2L_xL_y}{2 \pi H}, 
\label{eq:vin}
\end{equation}
and 
\begin{equation}
M_A = \frac{v_{in}}{v_A}
    = \frac{(v_{foot}B_{foot})^3(L_xL_y)^2}{(2 \pi H)^2}
      \sqrt{4 \pi m_p n_0}, 
\label{eq:ma}
\end{equation}
where the parameters in the right hand sides are observable.

\section{NUMERICAL SIMULATION OF A FLARE LOOP} \label{sec:nsim}

In order to derive the energy release rate $H$ from SXT observations, 
we have carried out a set of numerical simulations of a flare loop 
with different loop length and energy input. The effects of 
Spitzer type thermal conduction, optically thin radiative cooling, 
and chromospheric evaporation are included.  
From the results of numerical simulations we calculated the 
expected count rates for specific filters of the SXT, 
and then re-calculated the average temperature, emission measure, 
and hence thermal energy by using the filter ratio method \citep{har92}, 
as we usually do when we analyze the real observational data. 
By doing this we can calculate the ratio $\alpha$ of $dE_{th}/dt$ obtained 
from the SXT data to the actual energy release rate $H$ for various 
parameters; $\alpha = (dE_{th}/dt)/H$. 
Then $\alpha$ is used to calculate $H$ 
from SXT data in section 4. In this paper we consider only the 
combination of Al 12 $\mu$m (Al12) filter and Be 119 $\mu$m (Be) 
filter. This pair of Al12 and Be filters has been frequently used in 
flare observations.

\subsection{Basic Assumptions and Equations} \label{subsec:baeq}
\begin{figure}
\epsscale{.80}
\plotone{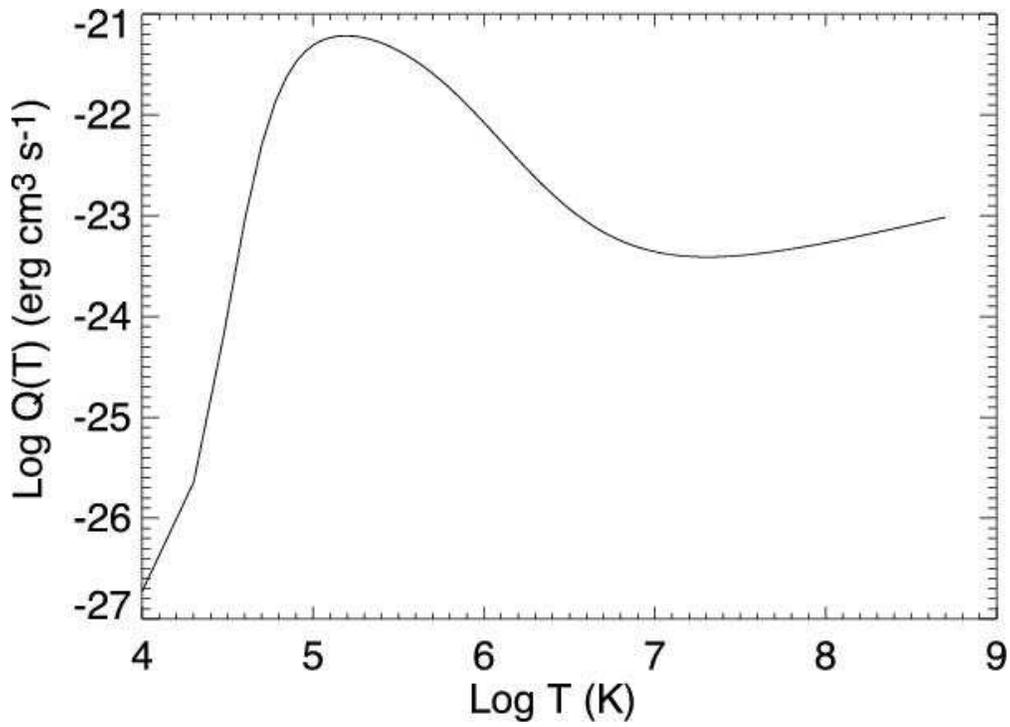}
\caption{Radiative loss function $Q(T)$.
\label{fig:radlosf}}
\end{figure}

As a simple model of a flare loop, we consider a one-dimensional 
magnetic loop with semi-circular shape and constant cross section. 
The basic equations are: 
\begin{eqnarray}
\label{eq:mass}
\frac{\partial \rho}{\partial t} + \frac{\partial}{\partial s}(\rho v) &=& 0, \\
\label{eq:momentum}
\frac{\partial \rho v}{\partial t} + \frac{\partial}{\partial s}(\rho v^2)&=&
      -\rho g_{\parallel} - \frac{\partial P}{\partial s}, \\
\label{eq:energy}
\frac{\partial e}{\partial t} + \frac{\partial}{\partial s}[(e + P)v] &=& 
      \frac{\partial}{\partial s}\Big( \kappa_{\parallel} \frac{\partial T}{\partial s
} 
\Big) 
      -\rho g_{\parallel}v - R + h, 
\end{eqnarray}
where
\begin{equation}
P = nk_BT, \quad e = \frac{1}{2}\rho v^2 + \frac{P}{\gamma -1}.
\end{equation}
Here $s$ is the distance along the loop from the photosphere. 
$\gamma = 5/3$ is the ratio of the specific heats. 
$g_{\parallel}(s)=g_0\cos (\pi s/2L)$ is the acceleration 
by the gravity along the loop; L is the 
distance from photosphere to the top of the loop. 
The radiative loss function $R=n^2Q(T)$ is chosen to approximate the form
for a optically thin plasma with normal solar abundances; 
$Q(T)$ is shown in figure \ref{fig:radlosf}. 
The thermal conductivity is taken to be of the classical Spitzer form 
\citep{spi56}, that is, $\kappa_{\parallel} = \kappa_0 T^{5/2}$ 
erg s$^{-1}$ cm$^{-1}$ K$^{-1}$ and $\kappa_0 = 10^{-6}$ in cgs units. 
The resulting heat flow is assumed not to be flux-limited. 
The heating term $h$ is divided into the static heating $h_s(s)$ 
and the flare heating $h_f(s,t)$. 
The static heating term $h_s(s)$ is assumed to balance the 
radiative losses in the initial condition at each point.
The second term $h_f(s,t)$ is the flare heating, which is described 
later.
The other parameters have their usual meanings.

The basic equations are numerically solved as an initial-boundary 
value problem. The hydrodynamic parts are solved by modified 
Lax-Wendroff method \citep{rub67} and the heat conduction part 
is solved by the successive overrelaxation method 
(e.g., \citealt{hir88}). 
The code has been extensively used for the modeling 
of solar flares \citep{hor97}, X-ray jets \citep{shm01} and 
protostellar flares \citep{iso03}. The detail of the numerical 
algorithms is described in the Appendix of \citet{hor97}.

\subsection{Initial and Boundary Conditions} 
The initial condition is hydrostatic atmosphere consisting of isothermal 
photosphere/chromosphere and hot corona. The temperature 
distribution in the initial condition is given by 
\begin{equation}
T=  T_{pho} + \frac{T_{cor} - T_{pho}}{2} \{ \tanh 
  \Big( \frac{s - s_{tr}}{w_{tr}} \Big)+1 \},  
\label{eq:inite}
\end{equation}
where $T_{pho}=10^4$ K, $T_{cor}= 2 \times 10^6$ K,  
$s_{tr} = 2.8 \times 10^8$ cm, and $w_{tr} = 10^7$ cm.  
The initial density and pressure distributions are calculated 
from the temperature and gravity distributions by solving 
hydrostatic equation. 

Since we assume that the static heating balances only with the radiative 
loss, the initial condition is not in strict energy 
equilibrium. Therefore, even without the flare heating 
slight decrease of temperature just above the 
transition region ($s \sim s_{tr}$) and 
weak evaporation of the chromosphere spontaneously occur 
due to the conductive heat flux from the hot corona to the cold 
chromosphere. However, they have negligible effects on the 
result of simulation because the temperature increase and resultant 
evaporation by the flare are much stronger.
The grid size $\Delta s$ smoothly decreases 
from $3.2 \times 10^6$ cm at the photosphere to 
$2.0 \times 10^7$ cm around the transition region. 
In the corona, $\Delta s$ increases from 
$2.0 \times 10^6$ cm to $1.0 \times 10^7$ cm. 

We assume symmetry about the loop top and solve only the half 
of the semi-circular loop, i.e., $0 \le s \le L$. Mirror boundary 
condition of the following form is imposed both at $s=0$ and 
$s=L$. 
Since our code requires that values of $\rho$, $\rho v$, 
and $P$ are specified on one grid point at each boundary, we set 
$\rho_1 = \rho_2$, $\rho_1 v_1 = \rho_2 v_2$, and 
$P_1 = P_2$ for the bottom boundary and 
$\rho_N = \rho_{N-1}$, $\rho_N v_N = \rho_{N-1} v_{N-1}$, and 
$P_N = P_{N-1}$ for the top boundary. Here the subscripts 
denote the position of the grid (i.e., $\rho_i$ is the density 
on the {\it i}th grid point), and 
$N$ is the total grid number. 
The bottom boundary condition does not affect the result 
of simulation because the gas and energy density near 
the bottom boundary is much larger than the upper atmosphere 
and therefore the bottom boundary is not perturbed significantly 
by the flare heating. We chose the mirror boundary because 
it is mathematically compatible and the total mass and energy 
in the simulation domain are conserved.

\subsection{Flare Heating}
The flare heating function $h_f(s,t)$ is of a form of spatially 
Gaussian, 
\begin{equation}
h_f(s,t)= \frac{q(t)}{\sqrt{2\pi}\sigma} 
          \exp [-\frac{(s-s_{flr})^2}{2\sigma^2}] \,
          \mathrm{(erg} \, \mathrm{cm}^{-3} \, 
          \mathrm{s}^{-1}\mathrm{)}. 
\label{eq:flare}
\end{equation}
We adopt $\sigma = 2.0 \times 10^8$ cm and $s_{flr} = L$, 
i.e., the flare energy is deposited at the loop top. 
The total heat flux for the loop $q(t)$ is set to be 
\begin{equation}
q(t)=\frac{q_f(1-\tanh \{(t-t_f)/w_{tf}\})}{2}. 
\end{equation} 
We adopt $t_f=360$ s and $w_{tf}=1.7$s, 
because the duration of the impulsive phase is about 6 minutes 
for all the flares analyzed in this paper. 
We also performed simulations with uniform heating and footpoint 
heating to see how the result varies if different form 
of flare heating is used. The comparison of the different 
heating models is presented in the Appendix.

\subsection{Results of Numerical Simulations}\label{subsec:result}
We performed the simulations with different loop length $L$ and 
heat flux $q_f$; $L=2.0,\, 3.0, \,3.3$, and $4.0 \times 10^9 \, \rm{cm}$ 
and 
$q_f = 0.4, \, 0.8, \, 4, \, 8, \, 40, \, 80$, and $243 \, 
\times 10^8 \, \rm{erg} \, \rm{s}^{-1} \, \rm{cm}^{-2}$. 
The other parameters were kept constant.
In the following we show the result of the run with 
$L=3.3 \times 10^9 \, \rm{cm}$ and 
$q_f=8.0 \times 10^8 \, \rm{erg} \, \rm{s}^{-1} \, \rm{cm}^{-2}$.

\begin{figure}
\epsscale{.80}
\plotone{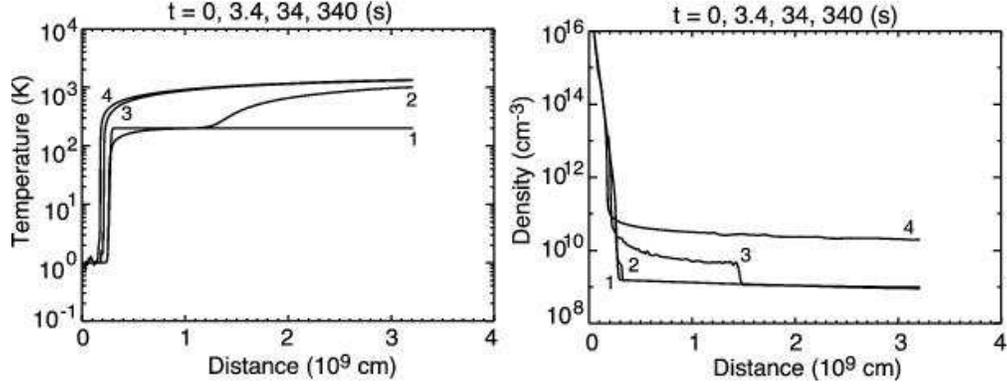}
\caption{Temperature and density distribution of the 
simulation result with typical parameters (see the text). 
The labels 1, 2, 3, and 4 in the figure indicate the time; 
$t=$ 0, 3.4, 34, and 340 s, respectively. 
\label{fig:tedeplots}}
\end{figure}

Figure \ref{fig:tedeplots} shows the temperature and density 
distributions at $t=$ 0, 3.4, 34, and 340 s. 
Since the flare energy is injected at the loop top, the 
temperature at the loop top rises rapidly, and the energy 
is transported by heat conduction. The conduction front is 
seen near $s=1.2 \times 10^9$ cm at $t=3.4$. When the conduction 
front reaches to the chromosphere, the cold and dense plasma 
is heated to the flare temperature, and the resultant high pressure 
drives the upward flows, i.e., chromospheric evaporation ($t=34$ s). 
The evaporation plasma fills the flare loop and increases the 
soft X-ray emission ($t=340$ s). 

\begin{figure}
\epsscale{.80}
\plotone{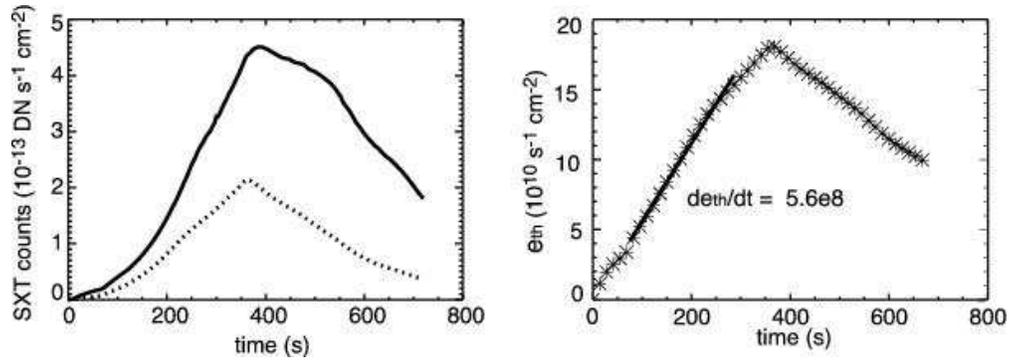}
\caption{Left: Synthesized light curves. The solid and dashed 
lines are the light curves of Thick Al. and Be filters, respectively. 
They must be multiplied by $3.2 \times 10^{16}$ to obtain the value in 
DN s$^{-1}$ pixel$^{-1}$ for full resolution image.
Right: Temporal variation of $e_{th}$, the thermal energy per 
unit area derived by the SXT filter ratio method. The solid line 
is the least square fitting of the impulsive phase.
\label{fig:sxt_eth}}
\end{figure}

From the temperature and density distributions in the result 
of simulation, the counts of signal that will be detected by SXT 
can be calculated by
\begin{equation}
I_j = \sum_{i=1}^N f_j(T_i)n_i^2\Delta s_i, 
\label{eq:sxtint}
\end{equation}
where $f_j$ is the response function of the filter $j$, 
$I_j$ is the count rate of SXT with filter $j$, 
$i$ denotes the grid number, and $\Delta s_i$ is the 
spacing of the $i$th grid. 
The left panel of figure \ref{fig:sxt_eth} shows 
the synthesized SXT count rates 
for Al12 filter ($I_{Al}$: solid line) and 
for Be filter ($I_{Be}$: dashed line). 
Note that the unit is DN (data number) per second per unit 
area, where the area is that on the solar surface. 
Since one pixel of the full resolution image of SXT corresponds 
to $2.45$ arcsec $\approx 1.8 \times 10^8$ cm on the solar surface,  
to obtain the count rate in DN s$^{-1}$ pixel$^{-1}$,  
the values in figure 4 must be 
multiplied by $3.2 \times 10^{16}$. 

From $I_{Al}$ and $I_{Be}$, the average temperature $T$ and emission 
measure $\epsilon_L=n^2L$ in the flare loop can be calculated 
by the filter ratio method (Hara et al. 1992). 
Here $\epsilon_L$ is the emission measure of the 
loop with a cross section of 1 cm$^2$. The thermal 
energy of the loop is given by 
\begin{equation}
e_{th}=\frac{2nk_BT}{\gamma -1}L
      =3\sqrt{\epsilon_LL}k_BT \,
          \mathrm{(erg} \, \mathrm{cm}^{-2}\mathrm{)}. 
\end{equation}
The total thermal energy $E_{th}$ is then given by 
$E_{th}=e_{th}S$, where $S$ is the cross section of the 
flare loop(s) which is arbitrary in the simulation.
Finally we obtain the ratio $\alpha$; 
\begin{equation}
\alpha = \frac{dE_{th}/dt}{H} = \frac{de_{th}/dt}{q_f}, 
\label{eq:alpha}
\end{equation} 
where  $q_f$ is the input heat flux of the flare.  
Note that the thermal energy calculated in this way is that of 
the high temperature plasma, because the SXT {\it sees} 
only the hot coronal plasma ($T \ge 2 \times 10^6$ K). 
The temporal variation of $e_{th}$ is shown in the right panel 
of figure \ref{fig:sxt_eth}. We defined the impulsive phase 
as $0.2t_{flr} \le 0.8t_{flr}$  and calculated the average gradient 
of the $e_{th}(t)$ in the impulsive phase by the least square fitting. 
The solid line in the right panel of figure \ref{fig:sxt_eth} 
shows the result of least square fitting that yields 
$de_{th}/dt = 5.6 \times 10^8$ erg s$^{-1}$ cm$^{-2}$. 
Since the input heat flux in this simulation is 
$8.0 \times 10^8$ erg s$^{-1}$ cm$^{-2}$, 
$\alpha =  0.7$ in this case. 

\begin{figure}
\epsscale{.80}
\plotone{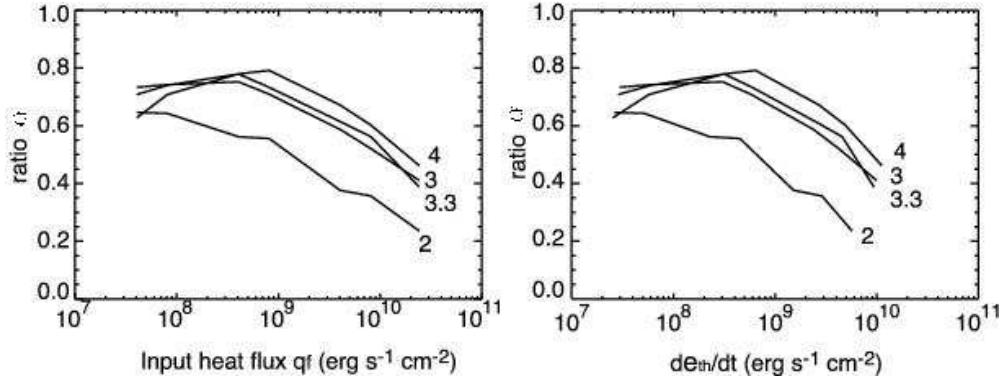}
\caption{The ratio $\alpha= (de_{th}/dt)/q_f$ plotted against 
$q_f$ (left) and $de_{th}/dt$ (right).
The numbers in the figure indicate the loop length ($10^9$ cm).
\label{fig:ratios}}
\end{figure}

We calculated $\alpha$ 
for all the other parameters by the same procedure. 
The result is shown in figure \ref{fig:ratios}. 
The left and right panels plot $\alpha$ against 
input heat flux and $de_{th}/dt$, respectively. 
The numbers in the figure indicate the loop length. 
Although there are some exceptions, basically larger 
heat flux or shorter loop length result in smaller $\alpha$. 
We interpret this tendency as follows. When the heat flux 
is larger or the loop length is shorter, the conductive 
flux into the chromosphere is larger, and hence the conductive 
cooling is more effective. Moreover, larger conductive flux leads to 
stronger chromospheric evaporation and larger density 
in the corona. Therefore radiative cooling is also effective in 
such cases. 
In the following analyses we use $\alpha$ 
to calculate the real energy release rate $H$ 
from $dE_{th}/dt$ obtained from SXT images.

\section{RECONNECTION RATE} 

Using the method described in section \ref{sec:howto} and 
the result of the simulation presented in section \ref{sec:nsim}, 
we have analyzed three flares:
an X2.3 class flare on 2000 November 24, 
an M3.7 class flare on 2000 July 14, 
and a C8.9 class flare on 2000 November 16. 
These flares were selected because 
(1) they show relatively symmetric expansion of two ribbons 
   in TRACE 1600 \AA \, image, 
(2) the SXT data taken with Al12 and Be filters are available 
   for the impulsive phases, 
(3) magnetograms from SOHO/MDI are available, and 
(4) the size ($\sim 4 \times 10^9$ cm) and the duration of the 
   impulsive phase ($\sim 6$ min) are similar. 

\subsection{X2.3 Flare on 2000 November 24}
This flare is one of the three X-class homologous flares 
occurred on 2000 November 24 in NOAA 9236. 
This active region produced many homologous flares and coronal 
mass ejections and has been extensively studied (e.g., 
\citealt{nit01}; \citealt{zha02}; \citealt{zha03}; \citealt{moo03}; 
\citealt{tak04}) 
A quantitative analysis of the ribbon separation and hard 
X-ray emission  of the homologous flares was presented by \citet{tak04}. 

\begin{figure}
\epsscale{.80}
\plotone{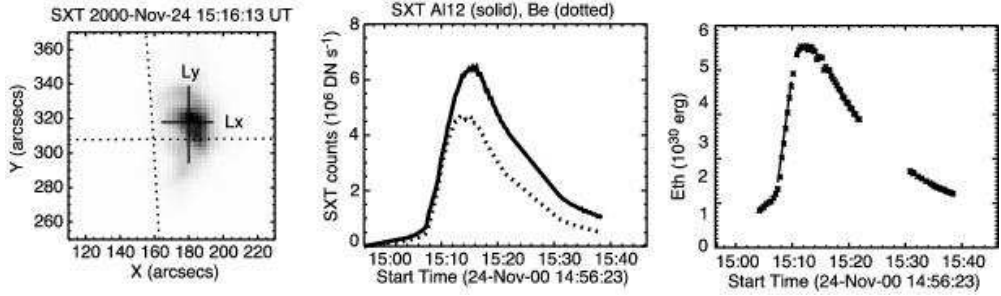}
\caption{Left: SXT image of the X2.3 flare on 2000 Nov. 24. 
The image is negative. The solid lines indicate $L_x$ (shorter line)
 and $L_y$ (longer line) measured by eyes. 
Middle: Light curves of the Al 12 filter images (solid line) and 
Be filter images (dotted line). 
Right: Temporal variation of thermal energy $E_{th}$. The solid line 
indicates the least square fitting of the impulsive phase, that 
yields $dE_{th}/dt=1.6 \times 10^{28}$ erg s$^{-1}$. 
\label{fig:sxt001124}}
\end{figure}

The SXT image near the peak is shown in figure \ref{fig:sxt001124} 
(left panel). The solid lines indicate $L_x=2.2 \times 10^9$ cm 
and $L_y=3.5 \times 10^9$ cm measured by eyes. 
The effect of projection on the plain of sky is corrected. 
Since the X-ray arcade grows with time and 
the arcade and flare ribbons are not exactly symmetric, 
the measured $L_x$ and $L_y$ are rough estimates. 
In this paper we consider the spatial and temporal 
average of the reconnection rate, so we use above values of 
$L_x$ and $L_y$ as the representative values to calculate 
the reconnection rate. In order to study the spatial and 
temporal variation of the reconnection rate, the size and 
other parameters must be measured more carefully considering  
the complicated and asymmetric structure. 

The middle panel of figure \ref{fig:sxt001124} shows the temporal 
variation of the intensity integrated over the flare arcade. 
The solid and dotted lines are for Al12 
and Be filter images, respectively. 
Since the SXT instrument cannot obtain two images with 
different filters simultaneously, we calculate the intensity 
of Al12 filter images at the same time of Be filter images 
by linear interpolation, and then calculate the temperature $T$ 
and the volume emission measure $\epsilon_V=n^2V$ by the filter ratio 
method. We assume that the line of light length of the arcade 
($\sim$ height of the arcade) is equals to $L_x$, the 
distance between the footpoints. This assumption is justified 
because we empirically know from the observations of limb flares 
that the height of a flare arcade is approximately equals to the 
distance between its footpoints. We also assume the volume filling 
factor $f=0.1$; $V=L_x^2L_yf$. This is also from our empirical 
knowledge that inner part of a flare arcade usually looks like 
a void. 
With these assumptions the thermal energy of the flare arcade is 
given by 
\begin{equation}
E_{th}=3nk_BTV = 3k_BT\sqrt{\epsilon_V V}.
\label{equation:eth} 
\end{equation}

The right panel of figure \ref{fig:sxt001124} shows the 
temporal variation of $E_{th}$. The solid line is the 
least square fitting in the impulsive phase. 
From the gradient of the solid line we obtain 
$dE_{th}/dt = 1.6 \times 10^{28}$ erg s$^{-1}$. 

We use the result of numerical simulation to calculate 
the real energy release rate $H$ from $dE_{th}/dt$. 
In order to compare with the one-dimensional numerical simulation, 
we need to know the energy release rate per unit area $de_{th}/dt$. 
To obtain this, we simply divide total energy release rate 
$dE_{th}/dt$ by the apparent area of the flare arcade $L_xL_y$; 
$de_{th}/dt= (dE_{th}/dt)/(L_xL_y)
 = 2.0 \times 10^{9}$ erg s$^{-1}$ cm$^{-2}$. 
From the assumed geometry we estimate that the loop length 
(distance from footpoint to looptop) is about 
$1.3 L_x \approx 3 \times 10^9$ cm. 
Then from figure \ref{fig:ratios} we see that 
$\alpha = 0.6$ for this flare. 
Thus we obtain 
$H=(dE_{th}/dt)/\alpha = 2.7 \times 10^{28}$ erg s$^{-1}$. 
These parameters are summarized in table 1. 

\begin{figure}
\epsscale{.80}
\plotone{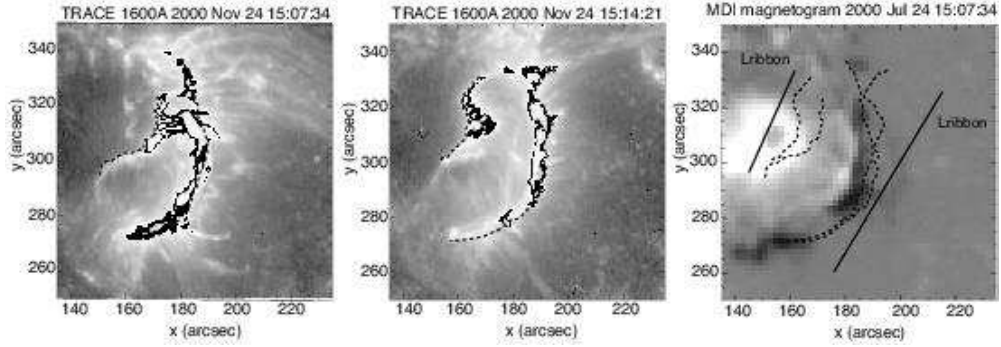}
\caption{Left and middle: TRACE 1600 \AA \, images of the X2.3 flare 
at $t_1=$15:07:34 and $t_2=$15:14:21. The dashed lines indicate the outer 
edge of the flare ribbons at each time. 
Right: MDI magnetogram of the same field of view. The dashed lines 
indicate the locations 
of the outer edges of the ribbons both at $t=t_1$ and $t_2$. 
The solid lines in the left panel indicate the length of the ribbon 
$L_{ribbon}$ for the east (left) and west (right) ribbons.  
\label{fig:trace001124}}
\end{figure}

Next step is to measure the separation velocity $v_{foot}$ 
and field strength $B_{foot}$ of the flare ribbons. 
Although the three flares analyzed 
in this paper have relatively symmetric and simple ribbons, 
$v_{foot}$ and $B_{foot}$ are different at the different 
position on the ribbons. Since we are interested in the 
average values, $v_{foot}$ and $B_{foot}$ are measured 
by following procedure. 
(1) Two TRACE 1600 \AA \, images, 
one near the beginning of the impulsive phase and 
one just before the end of the impulsive phase, are selected.  
(2) The MDI magnetogram at the nearest time is selected and 
    coaligned with the TRACE 1600 \AA \, images. 
   The coalignment is done by taking cross-correlation of the 
   TRACE white-light image and the MDI continuum image. 
(3) The outer edges of the ribbons in the TRACE 1600 \AA \, images 
   are defined by visual inspection of the images 
   (see figure \ref{fig:trace001124}). 
(4) The magnetic flux $\phi = \int B dA$, where 
    $A$ is the area swept by the outer edge of the flare ribbons, 
   is calculated from the coaligned magnetogram. 
    We assume that magnetic field is vertical at the photosphere, and 
    correct the effect of projection. The average field strength 
    $B_{foot}$ is given by $B_{foot}=\phi/A$.  
(5) The (average) separation velocity $v_{foot}$ is calculated by 
\begin{equation}
v_{foot}=\frac{A}{L_{ribbon}(t_2-t_1)}, 
\end{equation}
where $L_{ribbon}$ is the length of the ribbon, and $t_2$ and 
$t_1$ are the times of the TRACE images ($t_2 > t_1$). 
This procedure is applied to each ribbon separately. 
We use the average $B_{foot}$ and $v_{foot}$ of the two ribbons to 
calculate the reconnection rate. 

Figure \ref{fig:trace001124} shows the TRACE 1600 \AA \, images 
near the beginning ($t_1=$15:07:34 UT) and 
end ($t_2=$15:14:21 UT) of the impulsive phase, 
as well as the coaligned MDI magnetogram. 
The dashed lines indicate the outer edges of the ribbons. 
The magnetic flux $\phi$ are calculated by summing up 
the flux of all the pixels between the ribbon edges at $t=t_1$ and 
$t=t_2$. The solid lines beside the locations of the ribbons (dashed lines) 
indicate $L_{ribbon}$ for the east (left on the image) 
ribbon and the west (right on the image) ribbon.  

Table \ref{table:001124} shows $\phi, L_{ribbon}, B_{foot}$ 
and $v_{foot}$ for the east and west ribbons and their average values. 
Effect of the projection has been corrected. 
It should be noted that the values of the magnetic flux $\phi$
of the two ribbons are not equal. Such flux unbalance has also 
been reported by \citet{fle01}. 
Finally, the $B_{cr}$, $v_{in}$, $M_A$, and  $E_{cs}$ 
are calculated by equations (\ref{eq:vinb}), (\ref{eq:bc}), 
(\ref{eq:vin}), and (\ref{eq:ma}) using the average values of 
$B_{foot}$ and $v_{in}$ as well as $H, L_x, L_y,$ and $n_0$ derived 
from the SXT data. The results are shown in table \ref{table:result}. 
We obtained $B_c = 41$ G, $v_{in} = 1.3 \times 10^7$ cm s$^{-1}$, 
$M_A = 0.047$, and $E_{cs} = 539$ V m$^{-1}$. 

\subsection{M3.7 Flare on 2000 July 14}
\begin{figure}
\epsscale{.80}
\plotone{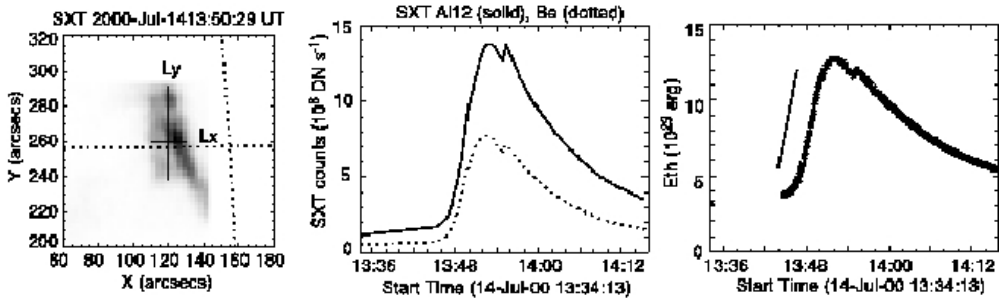}
\caption{The same figure as figure \ref{fig:sxt001124} for 
the M3.7 flare on 2000 Jul. 14. 
$dE_{th}/dt=3.5 \times 10^{27}$ erg s$^{-1}$.
\label{fig:sxt000714}}
\end{figure}

This flare occurred in NOAA 9077 at 13:45 UT on 2000 July 14, 
the same day as the famous 2000 Bastille Day flare but in 
the different active region. The SXT image, light curves, and 
the temporal variation of $E_{th}$ are shown in figure 
\ref{fig:sxt000714} in the same manner as figure \ref{fig:sxt001124}. 
Since the solid line showing the least square fitting of $E_{th}(t)$ 
is almost invisible, a line with the same slope ($=dE_{th}/dt$) is 
also shown in the figure. 

Following the same procedure as the 2000 Nov 24 flare, 
we obtained $L_x = 1.5 \times 10^9$ cm, $L_y = 4.2 \times 10^9$ cm, 
and $dE_{th}/dt = 3.5 \times 10^{27}$ erg s$^{-1}$. 
The ratio $\alpha$ is taken from figure \ref{fig:ratios}. 
The loop length is $1.3 L_x \approx 2 \times 10^9$ cm and the 
energy release rate per unit area is $(de_{th}/dt) 
\approx 5.6 \times 10^8$ erg s$^{-1}$ cm$^{-2}$, hence 
$\alpha = 0.55$ and 
$H = (dE_{th}/dt)/\alpha = 6.4 \times 10^{28}$ erg s$^{-1}$. 

\begin{figure}
\epsscale{.80}
\plotone{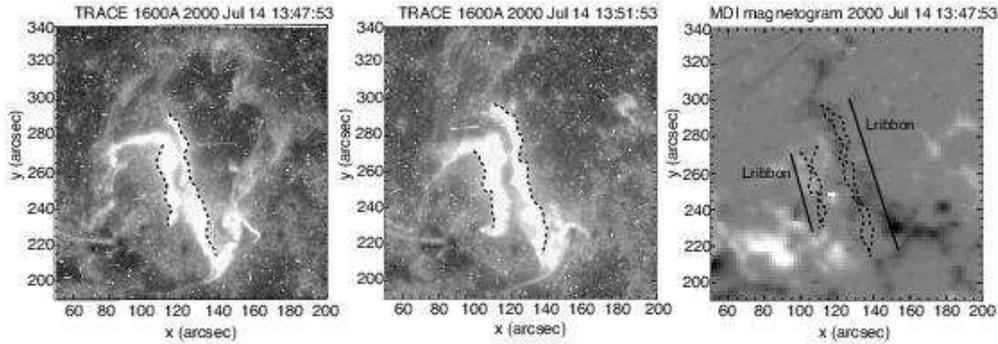}
\caption{The same figure as figure 7 for the M3.7 flare on 2000 Jul. 14. 
\label{fig:trace000714}}
\end{figure}

The TRACE 1600 \AA \, images and magnetogram of the flare are shown 
in figure \ref{fig:trace000714}, and the values of $\phi, B_{foot},$ 
and $v_{foot}$ are shown in table \ref{table:000714}. 
When we define the outer edge of the flare ribbons to calculate the 
magnetic flux, the upper part of the east (left) ribbon and the 
lower part of the west (right) ribbon that exhibit irregular shape 
are neglected because they seem to deviate from the standard 
two-dimensional geometry. 
Nevertheless the unbalance of $\phi$ is small. 
Finally we obtain $B_{cr}=44$ G, $v_{in} = 3.2 \times 10^6$ 
cm s$^{-1}$, $M_A = 0.015,$ and $E_{cs} = 143$ V m$^{-1}$. 
Other parameters are shown in table \ref{table:result}.

\subsection{C8.9 Flare on 2000 November 16}
\begin{figure}
\epsscale{.80}
\plotone{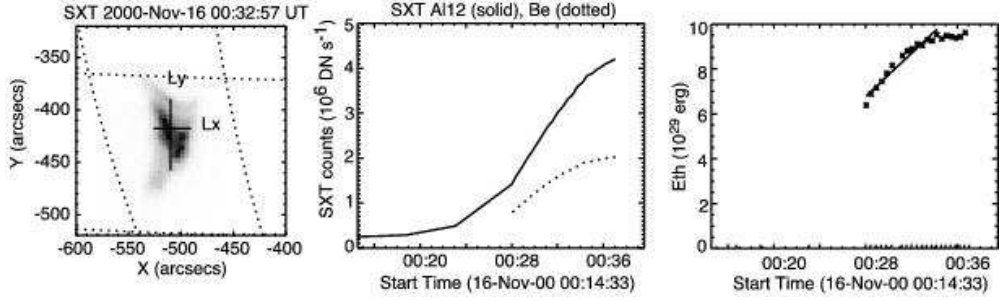}
\caption{The same figure as figure \ref{fig:sxt001124} for 
the C8.9 flare on 2000 Nov. 16. 
$dE_{th}/dt=1.2 \times 10^{27}$ erg s$^{-1}$.
\label{fig:sxt001116}}
\end{figure}

\begin{figure}
\epsscale{.80}
\plotone{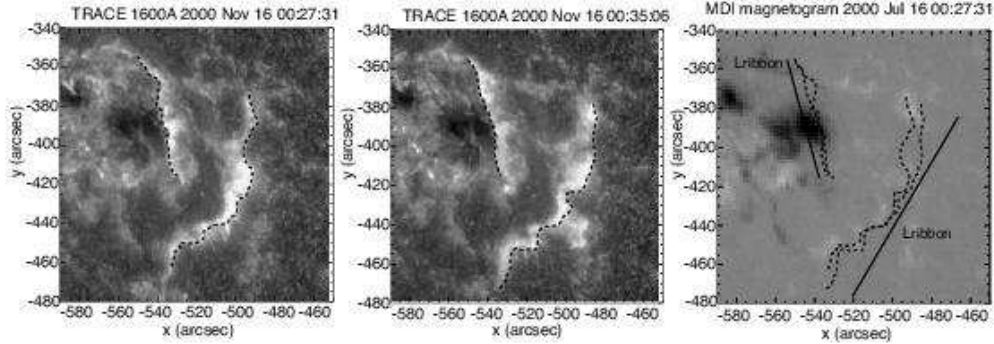}
\caption{The same figure as figure 7 for the C8.9 flare on 2000 Nov. 16. 
\label{fig:trace001116}}
\end{figure}

This flare occurred at 00:20 UT on 2000 November 16 in NOAA 9231. 
The SXT image, light curves, and the temporal variation of 
$E_{th}$ are shown in figure \ref{fig:sxt001116}, and the TRACE 
1600 \AA \, images and the MDI magnetogram are shown in figure  
\ref{fig:trace001116}, in the same manner as previous two flares. 
From the SXT data we obtain 
$L_x = 3.1 \times 10^9$ cm, $L_y= 5.2 \times 10^9$ cm,  
and $dE_{th}/dt = 1.2 \times 10^{27}$ erg s$^{-1}$. 
The loop length $1.3 L_x \approx 4.0 \times 10^9$ cm 
and the energy release rate per unit area 
$de_{th}/dt = 7.4 \times 10^7$ erg cm$^{-2}$, 
hence from figure \ref{fig:ratios} 
$\alpha = 0.65$ and $H=1.9 \times 10^{27}$ erg s$^{-1}$. 
Measurement of $\phi, B_{foot},$ and $v_{foot}$ are also done in 
the same manner. The results are shown in table \ref{table:001116}. 
Finally we obtain $B_{cr}=11$ (G), $v_{in} = 6.7 \times 10^6$ 
cm s$^{-1}$, $M_A = 0.071,$ and $E_{cs} = 70$ V m$^{-1}$.
Other parameters are shown in table \ref{table:result}. 

\subsection{Estimate of Uncertainty} \label{sec:error} 
It is difficult to estimate the uncertainties of the obtained values of 
$M_A, v_{in},$ and $B_{cr}$, because there are many factors that 
cause the uncertainty, though our method includes 
fewer assumptions than previous studies. 
Here we try to make a rough estimation of the uncertainty.   
First of all we should note that the obtained reconnection rate 
is the spatial and temporal 
average in the impulsive phase. Actually the reconnection 
process is often noisy and intermittent, which is clear from 
the highly variable, bursty light curves of nonthermal emissions 
such as hard X-ray and microwave. 
Careful examinations of the flare ribbons also suggest that the 
reconnection rate is noisy both in space and in time \citep{sab02, fle03}. 
Therefore each elementary process of reconnection in the flares 
can be faster than the average value obtained here. 

The thermal energy contents of the flare calculated in this study is 
consistent with previous works (e.g., Ohyama \& Shibata 1997, 1998; 
Emslie et al. 2004). However, we wish to measure the energy 
release rate $H$ as accurately as possible, because uncertainty in 
$H$ has significant influence on the calculated value of the 
reconnection rate.
The possible source of the uncertainty in the energy release rate $H$ is: 
(1) temperature and emission measure diagnostics by SXT, 
(2) error in the conversion from $dE_{th}/dt$ (measured from SXT data) 
    to $H$ using the results of the numerical simulations,  
and (3) measurement of the volume $V$. 
As for (1), the statistical error due to the photon noise is negligible. 
As for (2), $\alpha$  may depends on the 
detail of the simulation model such as the initial condition, 
location of the flare heating, etc. 
We performed simulations with different parameters and found 
that the detail of initial condition does not have significant 
consequence in the result when the total energy of the flare 
is much larger than the initial thermal energy contents of 
the loop. On the other hand, simulations with different 
heating locations indicate that if most of the 
flare energy is deposited deep in the chromosphere, 
$\alpha$ parameter may be significantly smaller. 
However, within the range of reasonable parameters
the uncertainty in the ratio $\alpha$ is about $\pm 20 \%$.  
The result of simulations with different heating 
locations is presented in the Appendix. 

Probably the largest uncertainty in $H$ comes from (3), 
the uncertainty of the volume. We made assumptions that the 
line of sight length of an arcade is equal to its footpoint distance 
$L_x$ and a volume filling factor of 0.1. We cannot tell precisely 
how good these assumptions are, but we empirically know that 
the assumption of the geometry (i.e., $L_z = L_x$)
is not a bad assumption, and 
probably the uncertainty is a factor of 2 at most. 
An alternative way to estimate the 
line of sight length may be to use the scaling law that relates 
the loop length (from apex to footpoint) to the rise time, 
decay time, and temperature of the flare \citep{met96}. 
We applied this scaling law by \citet{met96} to the flares 
analyzed in this paper and obtained the loop length of  
$1.3 \times 10^9$ cm, $1.3 \times 10^9$ cm, and 
$2.6 \times 10^9$ cm for the X2.3, M3.7, and C8.9 flares, 
respectively. These values are 1.5-2.2 times smaller that 
our assumption that the loop length is equal to $1.3L_x$. 
This also supports that the uncertainty of the line of sight 
length is about a factor of 2. 

The uncertainty of the volume filling factor is more difficult 
to address, but it seems that an uncertainty of a factor of 5 
is reasonable. 
The upper limit comes again from our empirical knowledge of 
the geometry of flare loops/arcades, and the lower limit 
comes from that if the filling factor is so small as 0.01, 
the number density of the plasma in the loops must be 
unreasonably large ($\sim 10^{13}$ cm$^{-3}$). 

Combining the uncertainties of the line of sight length and 
the filling factor, we conclude that the uncertainty of the volume 
is a factor of 10. From equation (\ref{equation:eth}) 
we see that the thermal energy is proportional to $\sqrt{V}$, hence 
the uncertainty in $E_{th}$ is about a factor of $\sqrt{10}$. 
Combining (1)--(3), we estimate that uncertainty in $H$ is about 
a factor of 4. 

About the measurement of $B_{foot}$ and $v_{foot}$, 
There must be errors that come from the alignment of 
the TRACE 1600 \AA \, ribbons and MDI magnetogram. 
The alignment of the TRACE images and MDI magnetograms  
has been done by taking the cross correlation of the white light 
images of TRACE and the continuum images of MDI. 
We found that the correlation between the two images was 
good in all the flares and estimated the error is about 
1 MDI pixel. This is consistent with the work by \citet{fle01} 
who used the same method for 
the coalignment of TRACE 1600 \AA \, images and MDI magnetogram.  
The magnetic flux $\phi$ measured by shifting the magnetogram 
by 1 pixel yield the uncertainty of a few $\%$. 

However, the unbalance of $\phi$ between the two ribbons 
indicates that there must be other source of uncertainty. 
As discussed by \citet{fle01} in detail, the uncertainty 
may come from:  
(1) the inclination of the photospheric field 
(2) not all the flux below the illuminated ribbons 
are involved in the reconnection event 
(3) uncertainty of the sensitivity of MDI magnetogram,  
    especially to the small scale, 
and so on. Although we do not know which factor is dominant, 
for the present purpose, it is enough to roughly estimate 
the uncertainty of the value of product $B_{foot}v_{foot}$. 
In the three flares analyzed in this paper, 
the largest difference of $E_{cs}=B_{foot}v_{foot}$ 
between two ribbons is about a factor of 2.4 
(2000 Nov 24 flare). If we adopt the values 
for the east and west ribbons as the upper and lower limits, 
the uncertainty of $B_{foot}v_{foot}$ may be considered 
to be a factor of $\sqrt{2.4} \approx 1.5$; 
\begin{equation}
\log_{10} (B_{foot}v_{foot}) = 
\log_{10} (B_{foot}v_{foot})_{\rm{average}} \pm \log_{10} 1.5 
\, \,  (\approx 0.2).
\end{equation}

From equations (\ref{eq:bc}) -- (\ref{eq:ma}), we see 
\begin{eqnarray*}
B_{cr} \propto H^1(B_{foot}v_{foot})^{-1} \\
v_{in} \propto H^{-1}(B_{foot}v_{foot})^2 \\
M_A    \propto H^{-2}(B_{foot}v_{foot})^{-3}. 
\end{eqnarray*}
Therefore, the estimated uncertainties are 
$\pm \log_{10} (4^1\times 1.5^1) \approx 0.8$ 
for $\log_{10} B_{cr}$, 
$\pm \log_{10} (4^1\times 1.5^2) \approx 1.0$ 
for $\log_{10} v_{in}$, and 
$\pm \log_{10} (4^2\times 1.5^3) \approx 1.7$ for 
$\log_{10} M_A$.  
 
For the uncertainty of the reconnection rate $M_A$, 
the uncertainty of the preflare density $n_0$ should be also 
taken into account. 
The largest uncertainty of $n_0$ comes from the uncertainty 
of the line of sight length $l$. However, since 
$v_A \propto n_0^{-1/2} \propto l^{1/4}$, the uncertainty 
of the preflare density can be neglected.

\section{DISCUSSION}
By considering the separation velocity of the flare ribbons 
and energy release rate, we have obtained dimensionless 
reconnection rate $M_A$ as well as the inflow speed $v_{in}$ 
and magnetic field strength $B_{cr}$ in the corona for the 
impulsive phases of three flares. 
The obtained values of reconnection rate (0.015-0.071) 
are consistent with the fast reconnection models 
(e.g., \citealt{pet64}). 
In other words we have confirmed that the motion of flare ribbons 
and the energetics of the flares are consistent with the standard 
scenario that the magnetic energy is released via fast magnetic 
reconnection. Furthermore, it seems that the large energy release rate $H$ 
and electric field $E_{cs}$ in the X-class flare are simply due to the 
large magnetic field strength and Alfv\'{e}n velocity in the corona. 

Previously several authors tried to estimate the 
non-dimensional reconnection rate from observational data 
(e.g., \citealt{der96}; \citealt{tsu96}; \citealt{ohy97}, 
\citeyear{ohy98}; \citealt{for00}) and obtained similar values. 
Our method has advantages to previous studies that it requires 
fewer assumptions and that the inflow velocity and magnetic field 
strength in the corona are also obtained  
(see discussion in Isobe et al. 2002). Therefore this method 
is also useful to estimate the coronal magnetic field strength. 

As mentioned in section \ref{sec:error}, the reconnection rate 
obtained above is the spatial and temporal average in the 
impulsive phase. Observations have shown that actually reconnection 
in solar flares is quite intermittent, both in time and in space
(e.g., Karlick\'{y} et al. 2005; Fletcher, Pollock, \& Potts 2003). 
Numerical simulations have also demonstrated that reconnection 
is intrinsically time dependent (e.g., Tanuma et al. 2001). 
Furthermore, based on the three-dimensional simulation 
Isobe et al. (2005) suggested that reconnection may be 
(intrinsically) intermittent not only in time but also in space.
Therefore, it is likely that in each elemental reconnection 
the reconnection rate is larger than the spatial and temporal 
average. In principle, the spatial and temporal variation of the 
reconnection rate can be determined by the same method 
used in this paper. 
In order to discuss the spatial and temporal variation of the 
non-dimensional reconnection rate, 
accurate alignment of the soft X-ray images 
(to derive the energy release rate), 
chromospheric images (such as $H\alpha$ or TRACE 1600 \AA), 
and the magnetogram are necessary. Although such analysis is 
possible using the present data, we hope that the data from 
the X-Ray Telescope (XRT) aboard Solar-B will be very suitable 
because of the high resolution and better temperature diagnostics. 

We obtained the inflow velocity $v_{in} = 30-130$ km s$^{-1}$. 
\citet{yok01} measured the velocity of the apparent motion of 
the bright structures in the EUV image of the inflow region 
and obtained $v_{in} \sim 5$ km s$^{-1}$. This value is smaller than 
the those obtained in this paper, probably because \citet{yok01} 
measured $v_{in}$ in the decay phase. Recently several studies related to 
\citet{yok01} have been done.  
Narukage \& Shibata (2005) found more examples of similar inflow events 
in SOHO/EIT data and obtained similar values of the reconnection rate.
\citet{cpf04} calculated the EUV image (FeXII 195 \AA \, emission line) 
from the result of numerical simulation and suggested the actual 
inflow velocity is larger than the velocity of the motion of 
bright pattern in the EUV image. On the other hand, 
\citet{nog05} analyzed the same event 
as \citet{yok01} using basically the same method as  
this paper and obtained the reconnection rate that is consistent 
with \citet{yok01}.  
More accurate measurement of $v_{in}$ should be done by detecting the 
Doppler shift by spectroscopy. It will be an important target of 
the EUV Imaging Spectrometer (EIS) of the Solar-B, the solar observing 
satellite to be launched in 2006. 

In order to make a constraint on the theoretical models 
of reconnection, it is valuable to examine the relation between 
the reconnection rate and the magnetic Reynolds number.
In the laboratory experiment, \citet{jih98} found that the 
measured reconnection rate in the 2D reconnection experiment 
can be explained by a generalized Sweet-Parker model that 
incorporates the compressibility and the effective 
resistivity. However, the magnetic Reynolds number of the reconnection 
experiments is less than $10^3$, hence it is not clear 
if the same model can be applied to the space and astrophysical 
reconnection where the magnetic Reynolds number is many orders of magnitude 
larger. 

The magnetic Reynolds numbers for the analyzed three flares are calculated 
using the vertical length of the current sheet $L_{z,cs}=L_x$ and 
the classical magnetic diffusivity given by \citep{pri82}
\begin{equation}
\eta=5.2 \times 10^{11} \ln \Lambda T^{-3/2} \; 
     \rm{cm}^2 \; \rm{s}^{-1}, 
\end{equation}
where for the Coulomb logarithm $\ln \Lambda \sim 20$ for the 
parameters considered here ($T \sim 10^7$ K, $n \sim 10^9$ cm$^{-3}$). 
To calculate $\eta$ we use the temperature derived from SXT data  
outside the flaring arcade ($T_0$), as was done to derive $n_0$. 

The values of $T_0$ and $R_m$ are shown in table \ref{table:result}. 
For the three flares analyzed here, there seems 
to be no dependence of the reconnection rate on the Magnetic Reynolds number. 
Of course the accuracy and the number of data is not enough to 
derive a conclusion on this issue. A statistical study of the 
reconnection rate of a large number of events will give us 
further insight into the physics to determine the reconnection rate.



\acknowledgments
We thank S. Nitta and T. Yokoyama for useful comments, 
and D. Shiota for drawing figure 1.
This work was supported by the Grant-in-Aid for the 21st Century 
COE "Center for Diversity and Universality in Physics" from the 
Ministry of Education, Culture, Sports, Science and Technology 
(MEXT) of Japan, and by Grant-in-Aid for Creative Scientific Research 
of the MEXT of Japan "The Basic Study of Space Weather Prediction"
(17GS0208, K. Shibata).
H.I. is supported by a Research Fellowship from the Japan 
Society for the Promotion of Science for Young Scientists.
We acknowledge the usage of the data from {\it Yohkoh}, SOHO, 
and TRACE.

\appendix

\section{Effect of Flare Heating Model}
\label{subsec:heat}
The form of the flare heating used in this paper 
assumes that all the released 
energy is injected at the loop top as the thermal energy 
and transported to the chromosphere by thermal 
conduction. However, if the significant fraction of 
the magnetic energy is converted to the non-thermal high energy 
particles during reconnection, the released energy 
is transported by the high energy particles and therefore 
the heating of the flare loop occurs in the transition region or 
chromosphere where the kinetic energy of the particles 
is thermalized through collision. 
Possibly, the factor $\alpha$ derived from the simulations 
may differ if the location of the flare heating is different. 
In order to test how the result depends on the location 
of the heating, we have performed simulations with 
uniform heating and footpoint heating. 

The uniform heating function is given by 
\begin{equation}
h_f(s,t)=\frac{q(t)}{\xi}
         \frac{\tanh [(s-z_{tr})/w_{tr}]+1}{2}, 
\end{equation}
where 
$\xi = \int_0^L \tanh [(s-z_{tr})/w_{tr}] ds$ is 
the normalization factor.
The footpoint heating function is given by 
\begin{equation}
h_f(s,t)= \frac{q(t)}{2\sqrt{2\pi}\sigma_d} 
          \exp [-\frac{(s-s_{tr}+d)^2}{2\sigma_d^2}] \,
          \mathrm{(erg} \, \mathrm{cm}^{-3} \, 
          \mathrm{s}^{-1}\mathrm{)}, 
\label{eq:flaref}
\end{equation}
where $d$ is the depth of the energy deposition from 
the transition region, and $\sigma_d=6 \times 10^7$ cm. 
We examined the cases with $d=0,2,4,$ and $6 \times 10^7$ cm. 
Compared with equation (\ref{eq:flare}), the factor of 1/2 in 
equation (\ref{eq:flaref}) is introduced so that the total heat 
flux in the loop is the same for the same value of $q(t)$. 

\begin{figure}
\epsscale{.80}
\plotone{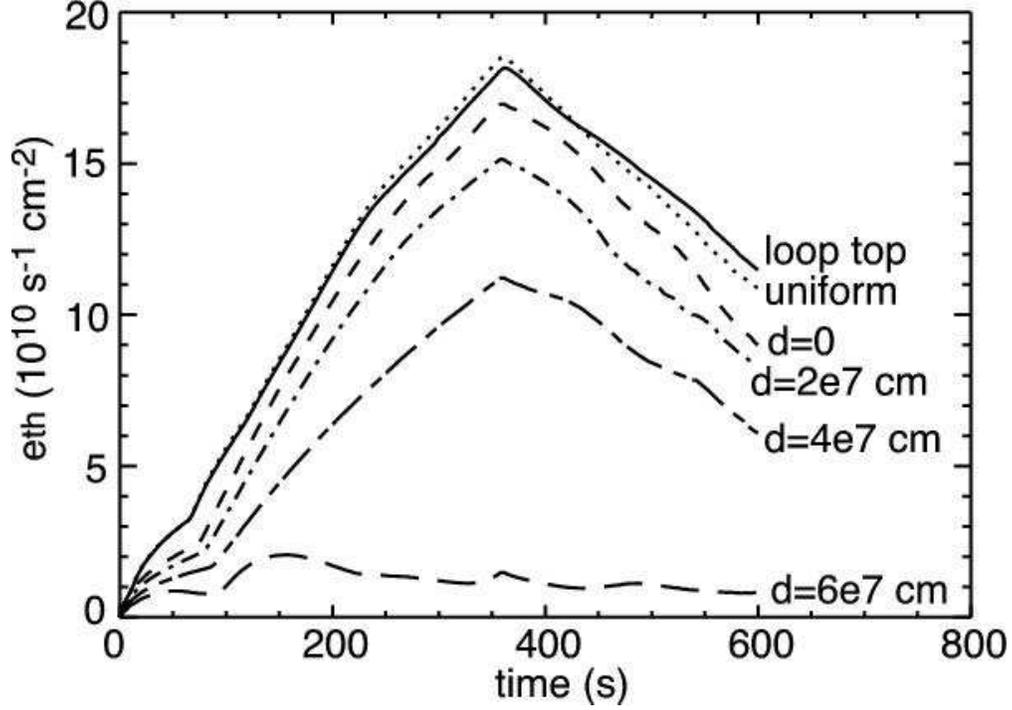}
\caption{Temporal variation of $e_{th}$ in the simulations with 
various flare heating fucntion. Plotted are the loop top heating, 
the uniform heating and the footpoint heating with $d=0,2,4$ and 
$6 \times 10^7$ cm, which are indicated in the figure. 
\label{fig:difheat}}
\end{figure}

Figure \ref{fig:difheat} shows the temporal variation of 
$e_{th}$, the total thermal energy in the loop derived from 
SXT filter ratio method. The heat flux $q_f$ is fixed to 
$8 \times 10^8$ erg s$^{-1}$ cm$^{-2}$ in all the cases. 
The evolution of the uniform heating case is almost identical 
to the loop top case which is described in section 
\ref{subsec:result}. This is because the temperature is almost 
kept constant in the corona due to the very effective 
thermal conduction. Therefore spatial distribution of the 
flare heating has little effect as long as it is in 
the corona. 

On the other hand, footpoint heating models result in 
smaller observed thermal energy. As shown in figure \ref{fig:difheat}, 
$e_{th}$ (and hence $de_{th}/dt$) is smaller when $d$ is larger. 
By examining the density and temperature profiles 
we found that this tendency is because the larger density in the 
location of flare heating results in larger radiative cooling. 
In the case of $d=0$ and $d= 2 \times 10^7$ cm the evolution of the 
flare loop is similar to that of loop top heating model, 
whereas in the case of $d=6 \times 10^7$ cm almost all the flare 
energy is immediately radiated in the chromosphere. 
The latter is probably an extreme case and not realistic, 
because in this case we expect strong hard X-ray emission 
by the precipitation of high energy particles and very 
weak soft X-ray emission, that is not consistent with 
the observed correlation with hard X-ray and soft X-ray 
intensities. Furthermore, by carefully examining figure 
\ref{fig:difheat} we find that difference in $de_{th}/dt$ 
(gradient of the curves in the rising phase) is smaller 
than that of the peak values of $e_{th}$, except for 
the $d=6 \times 10^7$ cm model.  
Therefore we conclude that the factor $\alpha$ 
derived in this paper assuming the loop top heating is 
quantitatively reasonable, with uncertainty of about $\pm 20 \%$. 
 
We should note that the large radiative cooling in the chromosphere 
may be due to the optically thin radiative cooling model, which is 
probaly not a good approximation in the actual solar chromosphere. 
More realistic modeling with proper treatment of non-LTE radiative 
transfer (e.g., Abbett \& Hawley 1999) is necessary for more acurate 
evaluation of the energy release rate. Furthermore, careful modeling 
of the chromospheric and transion region emissions is desirable to 
make spectroscopic diagnostics to determine what fraction of the 
released energy goes to the non-thermal particles and what goes 
to the thermal energy.




\clearpage



\begin{table}
\begin{center}
\caption{Parameters of the flares.\label{table:result}}
\begin{tabular}{crrr}
\tableline\tableline
Parameter    & 2000 Nov 24 & 2000 Jul 14& 2000 Nov 16 \\
\tableline
GOES class   & X2.3                 &  M3.7    & C8.9        \\
$H$ (erg s$^{-1})$        & 2.7e28  & 6.4e27   & 1.9e27        \\
$v_{foot}$ (cm s$^{-1})$  & 1.2e6   & 1.2e6    & 6.7e5    \\
$B_{foot} (G)$ &             449    & 117      & 106      \\
$L_x$ (cm) & 2.2e9 & 1.5e9    & 3.1e9    \\
$L_y$ (cm)                  & 3.5e9 & 4.2e9    & 5.2e9    \\
$n_0$ (cm$^{-3}$)           & 1.0e9 & 2.0e9    & 6.e8       \\
$T_0$ (K)                   & 1.0e7 & 8.0e6    & 8.e6           \\
$B_{cr}$ (G)  &               41    & 44       &  11   \\
$v_{in}$ (cm  s$^{-1})$ &    1.3e7  & 3.2e6    & 6.7e6  \\
$v_A$  (cm  s$^{-1}$) &      2.8e8  & 2.1e8    & 9.4e7 \\
$M_A$  &                     0.047 & 0.015    & 0.071        \\
$R_m$  &                     1.8e15 & 7.0e14   & 6.3e14 \\
$E_{cs}$ (V m$^{-1}$) &      539   &  143     &  70     \\
$\alpha $             &       0.6   &  0.55    & 0.65    \\
\tableline
\end{tabular}
\end{center}
\end{table}

\begin{table}
\caption{Magnetic flux $\phi$, ribbon length $L_{ribbons}$, 
magnetic field strength $B_{foot}$ and separation velocity $v_{foot}$ 
of the 2000 Nov. 24 flare. The effect of projection was corrected.
\label{table:001124}}
\begin{tabular}{cccc}
\tableline\tableline
Parameter  & east ribbon & west ribbon & average \\
\tableline
$\phi$ (Mx)  & 1.0e21 & 7.6e20 & 8.8e20  \\
$L_{ribbon}$ (cm)  & 3.1e9 &5.8e9 &4.5e9 \\
$B_{foot}$ (G) & 574 & 324 & 449 \\
$v_{foot}$ (cm s$^{-1}$)& 1.4e6 &1.0e6  &1.2e6  \\
\tableline
\end{tabular}
\end{table}

\begin{table}
\caption{Magnetic flux $\phi$, ribbon length $L_{ribbons}$, 
magnetic field strength $B_{foot}$ and separation velocity $v_{foot}$ 
of the 2000 Jul. 14 flare. The effect of projection was corrected.
\label{table:000714}}
\begin{tabular}{cccc}
\tableline\tableline
parameter  & east ribbon & west ribbon & average \\
\tableline
$\phi$ (Mx)  & 1.6e20 & 1.5e20 & 1.6e20  \\
$L_{ribbon}$ (cm)  & 3.4e9& 6.6e9& 5.0e9\\
$B_{foot}$ (G) & 144  & 91 & 117  \\
$v_{foot}$ (cm s$^{-1}$)  & 1.4e6 & 1.0e6 & 1.2e6 \\
\tableline
\end{tabular}
\end{table}

\begin{table}
\caption{Magnetic flux $\phi$, ribbon length $L_{ribbons}$, 
magnetic field strength $B_{foot}$ and separation velocity $v_{foot}$ 
of the 2000 Nov. 16 flare. The effect of projection was corrected.
\label{table:001116}}
\begin{tabular}{cccc}
\tableline\tableline
Parameter  & east ribbon & west ribbon & average \\
\tableline
$\phi$ (Mx)  & 1.4e20 &  2.2e20 & 1.8e20  \\
$L_{ribbon}$ (cm)  & 5.3e9& 7.6e9 & 6.4e9 \\
$B_{foot}$ (G) & 143 & 69  & 106 \\
$v_{foot}$ (cm s$^{-1}$) & 4.0e5 & 9.3e5  & 6.7e5 \\
\tableline
\end{tabular}
\end{table}






\end{document}